\documentclass{article}
\usepackage{times}
\usepackage{graphics}
\usepackage{epsfig}
\begin{document}
%\raggedright
%\twocolumn
%\setlength{\columnwidth}{16.4cm}
%\begin{minipage}[t][4.3cm][s]{17.7cm} {\sffamily
%\begin{center}
%\textbf{%%%%%%%%%%%%%%%%%%%%%%%%%%%%%%%%%%%%%%%%%%%%% TITLE (ALL
%%%%%%%%%%%%%%%%%%%%%%%%%%%%%%%%%%%%%%%%%%%%% UPPERCASE)
\title{Genetic selection of neutron star structure matching the X-ray
observations}
%\par%\medskip\par %%%%%%%%%%%%%%%%%%%%%%%%% AUTHORS (PRESENTING AUTHOR BOLD)
\author{Zden\v{e}k STUCHL\'IK\footnotemark[1], Petr
  \v{C}ERM\'AK\footnotemark[2], Gabriel
  T\"OR\"OK\footnotemark[1],\\ Martin URBANEC\footnotemark[1]
  and Pavel BAKALA\footnotemark[1]}
%\par%\medskip\par %%%%%%%%%%%%%%%%%%%%%%%%%%%%%%%%%%%%%%%%%%%%% AFFILIATIONS
%\normalsize
\maketitle
\footnotetext[1]{Institute of Physics, Faculty of Philosophy and Science,\\
  Silesian University in Opava, CZECH REPUBLIC}
\footnotetext[2]{Institute of Computer Science, Faculty of Philosophy and Science,\\
  Silesian University in Opava, CZECH REPUBLIC}

%\par\textsuperscript{2}%
%\textsl{Another Institution If Required, KAZAKHSTAN}
%\par\medskip\par %%%%%%%%%%%%%%%%%%%%%%%%%%%%%%%%%%%%%%%%%%%%
%ABSTRACT TEXT
%\end{center}
%\maketitle
%%\end{minipage}

%\begin{minipage}[t]{17.7cm}
%[10.6cm][s]{17.8cm}
%\vspace{0.5cm}
%\parbox[t]{8.2cm}{
%\begin{center}
%\abstract
%\end{center}
%\vspace{0.5cm}
\begin{abstract}
Assuming a resonant origin of the quasiperiodic oscillations
observed in the X-ray neutron star binary systems, we apply a
genetic algorithm method for selection of neutron star models. It
was suggested that pairs of kilo-hertz peaks in the X-ray Fourier
power density spectra of some neutron stars reflect a non-linear
resonance between two modes of accretion disk oscillations. We
investigate  this concept for a specific neutron star source. Each
neutron star model is characterized by the equation of state (EOS),
rotation frequency $\Omega$ and mass $M$. These determine the
spacetime structure governing geodesic motion and position dependent
radial and vertical epicyclic oscillations related to the stable
circular geodesics. Particular kinds of resonances (KR) between the
epicyclic frequencies, or the frequencies derived from them, can
take place at special positions assigned ambiguously to the
spacetime structure. The {pairs of resonant
  eigenfrequencies} relevant to those positions are therefore fully
given by KR,$M,~\Omega$, EOS  and can be compared to the
{observationally determined pairs of eigenfrequencies} in order to
eliminate the unsatisfactory sets (KR,$M,~\Omega$, EOS). For the
elimination we use the advanced genetic algorithm. Genetic algorithm
comes out from the method of natural selection when subjects with
the best adaptation to assigned conditions have most chances to
survive. The chosen genetic algorithm with sexual reproduction
contains one chromosome with restricted lifetime, uniform crossing
and genes of type 3/3/5. For encryption of physical description
(KR,$M,~\Omega$, EOS) into chromosome we used Gray code. As a
fitness function we use correspondence between the observed and
calculated pairs of eigenfrequencies.
\end{abstract}
%\keywords{black holes -- naked singularities -- X-ray variability -- theory --
%observations}

%\vspace{0.5cm}
%\begin{center}
\section{Introduction}

%\section{Introduction}%%%%%%%%%%%%%%%%%%%%%%%%%%%%%%%%%%%%%%%%%%%%%%%
Recently developed observational techniques provide good quality
data from observations of quasiperiodic oscillations (QPOs) in black
hole and neutron star sources \cite{Stu:2007}. It is shown that in
the case of some neutron star atoll sources (e.g. 4U1636-53
\cite{Bar}) the data can be well fitted by the so called
 multiresonant total precession model \cite{Stu:2007}. The fits give high precision values of the
 neutron star spacetime parameters;
in addition, in some cases the rotation
frequency of the neutron star is measured determined almost exactly
 from the QPO independent measurements. This enables
us to put some limits on equations of state (EOS) describing the
neutron star interior. Here we focus our attention of the EOS
given by the Skyrmion interaction that are very well tuned to the data
given by the nuclear physics \cite{Sto-etal:2003:PHYSR3:}. Using the
genetic algorithm method, which appears to be very fast and efficient,
we select the acceptable EOS from 27 types of the Skyrmion EOS selected by other methods
\cite{Sto-etal:2003:PHYSR3:}, concentrating on the source 4U
 1636-53. For the other five atoll sources, the method gives similar results.

\section{Fitting the QPO data}%%%%%%%%%%%%%%%%%%%%%%%%%%%%%%%%%%%%%%%
The results of recent studies of neutron star QPOs indicate that for a
given source the upper and lower QPO frequency can be traced through
the whole range of observed frequencies but the probability to detect
both QPOs simultaneously increases when the frequency ratio is close to
ratio of small natural numbers (namely 3/2, 4/3, 5/4 in the case of
atoll sources studies recently, see \cite{Stu:2007}). Therefore, the
multi--resonant orbital model based on the oscillation with Keplerian
($\nu_\mathrm K$) and epicyclic vertical ($\nu_\theta$) or radial
($\nu_\mathrm r$) frequencies was used to explain the observed
data \cite{Stu:2007}. They are calculated assuming the spacetime given
by the Hartle--Thorne metric
\cite{Har-Tho:APJ:1968,Chan-Mil}. The fitting procedure have shown that the
best results are obtained using the total precession model, where in
all the sources the upper frequency $\nu_\mathrm u=\nu_\mathrm K$ and
the lower frequency $\nu_\mathrm l=\nu_\mathrm T=\nu_\mathrm \theta -
\nu_\mathrm r$ \cite{Stu:2007}. We concentrate here on the case
of 4U 1636-53, when the mass and dimensionless spin of the neutron
star are fitted to values ($\chi^2 +1$ precision) in the range
\begin{eqnarray}
M&=&1.77 \pm 0.07,~~~~j=0.051 \pm 0.044, \\
M&=&1.84 \pm 0.07,~~~~j=0.101 \pm 0.044.
\end{eqnarray}
This interval of allowed values of $M$ and $j$ will be used to test the
EOS for the neutron star in the 4U 1636-53 source.
%}
%\hfill
%%\begin{minipage}[tr][12cm][s]{8.2cm}
%\parbox[t]{8.2cm} {

%}
%\end{minipage}

%\begin{minipage}[t][17.7cm][s]{17.7cm}
%\vspace{-0.5cm}
\section{The neutron star structure}%%%%%%%%%%%%%%%%%%%%%%%%%%%%%%
In neutron star, the strong gravity (i.e. Einsteins gravitational
equations) must be relevant being crucial for the structure
equations. The spacetime geometry is assumed to be stationary and
axisymmetric, the perturbative approach of Hartle and Thorne
\cite{Har-Tho:APJ:1968} is used. One start with the
spherically symmetric space time, given by Schwarzschild--like metric
\begin{eqnarray}
\mathrm{d} s^2&=&-\mathrm
e^\nu \mathrm{d}t^2+\left[1-2m(r)/r\right]^{-1}\mathrm{d}
r^2 \nonumber \\
&+& r^2\left(\mathrm d \theta^2 + \sin^2
  \theta (\mathrm d \phi-\omega \mathrm d t^2) \right].
\end{eqnarray}
To calculate structure of nonrotating, unperturbed star one has to
integrate TOV equation
\begin{equation}
\frac{\mathrm d P}{\mathrm d r}=-\frac{Gm(r)\rho}{r^2}\frac{\left(1+P/\rho c^2\right)
  \left[1+4\pi r^3P/m(r)c^2\right]}{1-2Gm(r)/rc^2} \label{TOV},
\end{equation}
where
\begin{equation}
m(r)=\int\limits_0^r 4\pi r^2 \rho \mathrm d r
\end{equation}

is the mass inside radius $r$. The integration is done outward from
the center (for given values of central energy density
$\rho_\mathrm{c}$) up
to surface (where pressure vanishes). One obtains the global properties of
non--rotating neutron star as its mass $M$, radius $R$ and the
internal characteristic profiles as metric coefficients, pressure, energy
density and number density of baryons expressed as functions of the
radial coordinate distance from center.

The rotational effect is, in the linear approximation, given by the
Hartle--Thorne metric \cite{Har-Tho:APJ:1968}.
\begin{eqnarray}
\mathrm{d} s^2&=&-\mathrm
e^\nu\left[1+2(h_0+h_2P_2)\right]\mathrm{d}t^2 \nonumber \\
&+&\frac{\left[1+2(m_0+m_2P_2)/(r-2M)\right]}{1-2M/r}\mathrm{d}
r^2 \nonumber \\
&+& r^2\left[1+2(v_2-h_2)P_2\right] \nonumber \\
&\times&\left[\mathrm d \theta^2 + \sin^2
  \theta (\mathrm d \phi-\omega \mathrm d t^2) \right],
\end{eqnarray}
where $P_2=P_2(\cos\theta)=(3\cos^2\theta-1)/2$ is the Legendre
polynomial of 2nd order $\omega$ is the angular velocity of local
inertial frame, which is related to star's angular velocity
$\Omega$ and $h_0,h_2,m_0,m_2$ are functions of $r$ and are all
proportional to $\Omega^2$.The angular velocity $\omega$ can be found
by solving equation
\begin{equation}
\frac{1}{r^4}\frac{\mathrm d}{\mathrm d r}\left(r^4 j \frac{\mathrm d
  \tilde{\omega}}{\mathrm d r} \right)+\frac{4}{r}\frac{\mathrm d
  j}{\mathrm d r}\tilde{\omega}=0 \label{omega}
\end{equation}
where
\begin{equation}
j(r)=\mathrm e^{-\nu(r)/2}[1-2m(r)/r]^{1/2}.
\end{equation}
One integrate equation (\ref{omega}) outward from center for arbitrarily
chosen $\tilde\omega_\mathrm c$ with boundary condition $\mathrm d
\omega/\mathrm d r=0$. At the surface one can calculate the angular
momentum $J$ and frequency of rotation $\Omega$ corresponding to
$\tilde\omega_\mathrm c$ from relations

\begin{eqnarray}
J&=&\frac{1}{6}R^4\left(\frac{\mathrm d \tilde\omega}{\mathrm d
  r}\right)_{r=R} \\
\Omega_\mathrm{new}&=&\tilde{\omega}(R)+\frac{2J}{R^3}.
\end{eqnarray}
We take frequency of rotation $\Omega$ as an
input parameter, thus after integration of eq. (\ref{omega}) we
rescale the  $\tilde\omega$ to obtain the right value of $\Omega$
\begin{equation}
\tilde{\omega}_\mathrm{new}(r)=\tilde{\omega}_\mathrm{old}(r)\frac{\Omega_\mathrm{new}}{\Omega_\mathrm{old}}.
\end{equation}
After this one should calculate mass and pressure perturbation factors
$m_0$ and $p_0$ from equations
\begin{eqnarray}
\frac{\mathrm d m_0}{\mathrm d r}&=& 4\pi r^2\frac{\mathrm d
  E}{\mathrm d P}(E+P)p_0 + \frac{1}{12}j^2r^4\left(\frac{\mathrm d
  \tilde\omega}{\mathrm d r}\right)^2 \nonumber \\
&-&\frac{1}{3}r^3\frac{\mathrm d
  j^2}{\mathrm d r}\tilde\omega^2,
\end{eqnarray}
%0&=&0\\
\begin{eqnarray}
\frac{\mathrm d p_0}{\mathrm d r} &=& -\frac{m_0(1+8\pi
  r^2P)}{(r-2m)^2} - \frac{4\pi(E+P)r^2}{r-2m}p_0 \nonumber \\
  &+&\frac{1}{12}\frac{r^4 j^2}{r-2m}\left(\frac{\mathrm d
  \tilde\omega}{\mathrm d r}\right)^2  +\frac{1}{3}\frac{\mathrm
  d}{\mathrm d r}\left(\frac{r^3j^2\tilde\omega^2}{r-2m} \right).
\end{eqnarray}
The mass of rotational object is then given by
\begin{equation}
M(R)=M_0(R)+m_0(R)+J^2/R^3.
\end{equation}
%%%%%%%%%%%%%%%%%%%%%%%%%%%%%%%%%%%%%%%%%%%%%%%%%%%%%%%%%%%%%%%%%%%%%%%%%%%%%%%%%%%%%%%%%%%%%%%%%%%%%%%%%!!!!!!!!!!

In the next approximation, the quadrupole moment $q$ of the star is
introduced. The fitting procedure show that $q \sim j^2$
\cite{Stu:2007}  where $j=J/M^2$. Therefore, the spacetime could be
considered quasi-Kerr and the Kerr metric and related formula of the
orbital motion could be used.
%%\end{minipage}
%%\begin{minipage}[tl][24cm][s]{8.2cm}

\section{Skyrmion interactions and realated EOS}%%%%%%%%%%%%%%%%%%%%%%%%%%%%%%%%%%%%
The effective Skyrmion interaction implies a variety of
parametrization in the framework of mean--field theory. All give
similar agreement with experimentally established nuclear ground
states at the saturation density $n_0$, but they imply varying
behaviour of both symmetric and asymmetric nuclear matter when
density grows (up to $3 n_0$).

The general form of effective Skyrme interaction implies total
binding energy of a nuclei as the integral of an energy density
functional $\mathcal{H}$, determined as a function of nine empirical
parameters $t_0, t_1, t_2, t_3, x_0, x_1, x_2, x_3$ and $\alpha$ in
the form

\begin{equation}
\mathcal{H}=\mathcal{K}+\mathcal{H}_0+\mathcal{H}_3+\mathcal{H_\mathrm{eff}},
\end{equation}
where the kinetic term $\mathcal{K}=\left(\hbar /2m \right)\tau$, is
given by the kinetic densities $\tau=\tau_\mathrm{n} +
\tau_\mathrm{p}$, with $\tau$ being the total
density,$\tau_\mathrm{n}$ ($\tau_\mathrm{p}$)is the  neutron (proton)
density. The other terms are

%}
%\end{minipage}

%\begin{minipage}[t][17.7cm][s]{17.7cm}
%\parbox[t]{8.2cm}{ \vspace{-0.8cm}
given by the relations
\footnotesize
\begin{eqnarray}
%\mathcal K&=& \frac{\hbar^2}{2m}\tau \\ \nonumber \\
\mathcal H_0&=& \frac{1}{4}t_0\left[(2+x_0)n^2 - (2x_0 +
  1)(n_p^2+n_n^2)\right] \\ \nonumber \\
\mathcal H_3 &=& \frac{1}{24}t_3n^\alpha \left[(2+x_3)n^2 - (2x_3 +
1)(n_p^2+n_n^2)\right]\\
\mathcal H_{eff} &=& {1}{8}\left[t_1(2+x_1)+t_2(2+x_2)\right]\tau_n\nonumber
\\ \nonumber \\
&+&\frac{1}{8}\left[t_2(2x_2+1)-t_1(2x_1+1)\right](\tau_p n_p+\tau_n n_n)
\end{eqnarray}
\normalsize

The pressure is then given by
\begin{equation}
P(n_\mathrm b,I)= n^2_\mathrm b \frac{\partial \varepsilon}{\partial n_\mathrm b},
\end{equation}
where $\varepsilon$ is the binding energy per particle and $I=(N-Z)/A$
denotes assymetry of nuclear matter.

In \cite{Sto-etal:2003:PHYSR3:}, 87 different Skyrme parametrization were
limited to 27, using limits implied by the spherically symmetric
models of  neutron
star models and by experimentally tested properties of nuclear matter. Here, we shall test
acceptability of nine of these 27 Skyrme parametrization to the limits put by QPO
measurements and using axisymmetric models in first approximations
with respect to the star rotation.

\section{Determination of neutron star structure using genetic algorithm}
Genetic algorithm (GA) appears from method of natural selection, when subjects with best
adaptation to assigned conditions have highest chance to survive \cite{Gold}. GA
takes into account the following natural mechanism - mutation
and lifetime limit restricting risk of degradation, which is kept
in local extreme from optimization viewpoint. GA has iteration
character. GA doesn't work with separate result in particular iterations,
but with population. In each iteration GA works with several
(generally a lot of results, standard value is hundreds) results,
which are included in the population trying to ensure appearance of still
better results via genetic operations with these results. Generally,
the GA scheme is given in the form
\begin{equation}
GA=(N,P,f,\Theta,\Omega,\Psi,\tau)
\end{equation}
where $P$ is population containing $N$ elements, $\Theta$ is parent
selection operator which selects $u$ elements from $P$. Evaluation for each
chromosome performed by the fitness function $f$.
\begin{equation}
f:S_i \leftarrow R,~  i=1,...N.
\end{equation}
Genetic operators included  in $\Omega$ are namely crossover operator
$\Omega_\mathrm C$, mutation operator $\Omega_\mathrm M$ and others
problem-oriented or implementation-oriented specific operators, which
all together generate $v$ offspring from $u$ parents. $\Psi$ is deletion
operator, which removes $v$ selected elements from actual population
$P(t)$. $v$ elements is add to new
%\vspace{-0.8cm}
population $P(t+1)$ after it, $\tau$ is
stop-criterion. Parent selection operator $\Theta$ and genetic
operators $\Omega$ have stochastic character, deletion operator
$\Psi$ is generally deterministic.

We have selected GA with sexual reproduction containing one chromosome
with restricted lifetime parameter, 5 iterations. The crossover
operator is selected with uniform crossing, using genes of type
3/3/5. We used the Gray code for encrypt parameters to
chromosomes, which is useful by reason of  bypassing so-called Hamming
barrier.

Chromosomes are compound of genes; each gene presents 1 bit value.
However, genes contain more than 1 bit (using the redundancy
encrypt). Bit values inside a gene are mapped onto outside value of
gene (0 or 1) via specific map function, in which border between 0
and 1 is not crisp, but there exists so-called ``shade zone'', where
value carried by the gene is determined randomly.
\cite{Cer:2006,Cer:2004}.We pressed number of members in each
generation to 400 (see Fig.1) and interate 40 generations. We define
two task of selection neutron star structure. The first one
determines $\rho_\mathrm c$, $\Omega$ with preset EOS where fitness
function f is $\chi^2$ \cite{Stu:2007}. It is important to say here,
that at present the fitness function is tabled. The second one
determines the EOS, $\rho_\mathrm c$, $\Omega$ with using fitness
function $\chi^2$. Both tasks put limits on the EOS under
consideration. Overcoming those limits implies removing the
corresponding chromosomes by setting value of fitness function to a
maximum value ($10^{200}$). For a given EOS allowed neutron star
structure is provided the GA described above and fitness function
$\chi^2$. The chromosome structure is given by two values, central
density and rotation frequency. Interval of central density we
choose from $0.4 \cdot 10^{15}$ to $3.3 \cdot 10^{15}~
\mathrm{g.cm}^{-3}$.

In the first task, partition of central density is set to 4096 (12
bits). Interval of rotation frequency is set from 50 to 8000 and
partition to 8192 (13 bits). Following table shows determination
of central density (in units of $10^{15}~ \mathrm{g.cm}^{-3}$) and
rotation frequency for given number of EOS.
%\begin{center}
%\bf Table 1. \rm
%\vspace{0.5cm}
\begin{table}\caption{Fitness function and corresponding
  parameters for all tested EOS, $\Omega \in (50;8000)$}
\centering
\begin{tabular}{l|c|c|c}
EOS type & $\rho_c$ & $\Omega$& Fitness \\ \hline
0-SkT5 & 2.1244 & 964.17 & 56.556 \\
1-SkO' & 1.5575 & 979.70 & 56.686 \\
2-SkO & 1.3987 & 711.85 & 56.601 \\
3-SLy4 & 1.4896 & 1373.71 & 57.292 \\
4-GS & 1.1225 & 743.89 & 57.223 \\
5-SkI2 & 1.1185 & 1241.72 & 57.688 \\
6-SkI5 & 0.9359 & 791.43 & 57.059 \\
7-SGI & 1.0170 & 986.49 & 57.028 \\
8-SV & 0.7889 & 590.55 & 56.393 \\
\end{tabular}
%\vspace{0.3cm}
\end{table}
%\vspace{0.5cm}
%\bf Table 1. \rm
%\end{center}
Because the resulting frequencies do not fit the observed rotation
frequency interval \cite{Lamb}, and also the fitness function has
too many of local minima with same value for different combination
of $\rho_\mathrm C,~\Omega$ ,we make determination of values
mentioned before, but we set range of rotation frequencies from
1827.21 to 1828.03 with partition 64 (6 bit).
%\end{minipage}
%\end{figure}
%}
%\end{minipage}
%\begin{minipage}[t][17.7cm][s]{17.7cm}

%\bf Table 2. \rm
%\begin{center}
%\begin{center}
\begin{table}\caption{Fitness function and corresponding
  parameters for all tested EOS, $\Omega \in (1827.21;1828.03)$}
\centering
\begin{tabular}{l|c|c|c}
EOS type & $\rho_c$ & $\Omega$& Fitness \\ \hline
0-SkT5 & 2.9978 & 1827.928 & 62.979 \\
1-SkO' & 1.7896 & 1827.453 & 57.542 \\
2-SkO & 1.6762 & 1827.210 & 57.899 \\
3-SLy4 & 1.5747 & 1827.300 & 57.611 \\
4-GS & 1.3084 & 1827.236 & 57.830 \\
5-SkI2 & 1.2174 & 1827.505 & 58.061 \\
6-SkI5 & 1.0803 & 1827.377 & 58.277 \\
7-SGI & 1.1146 & 1827.492 & 58.658 \\
8-SV& 0.8858 & 1827.428 & 58.45 \\
\end{tabular}
\end{table}
%\end{center}
%\end{center}
Clearly, for selecting the most convenient EOS we need referenced
$\chi^2$ value as global minimum $\chi^2$ over all EOS. Thus we make
second task, determination of the best resulting neutron star
structure over all used EOS. The corresponding EOS value is given by
zero based EOS number (4 bits).

%\bf Table 3. \rm
%\begin{center}
%\vspace{0.5cm} \small{
\begin{table}\caption{Global minima for $\Omega \in
(1827.21;1828.03)$}\centering
\begin{tabular}{c|c|c|c|c|c}
$\Omega_\mathrm{min}$ & $\Omega_\mathrm{max}$ & Fitness & EOS
  &$\rho_c$ & $\Omega$ \\ \hline
%50 & 8000 & 56.390 & 3& 1.38811 & 739 \\
1827.21 & 1828.03 & 57.542 & 1 &1.79021 & 1827.66 \\
\end{tabular}
%\vspace{0.3cm}
\end{table}
%\end{center}

\begin{figure}[t]\centering
%%\begin{minipage}{0.5\hsize}
%\resizebox{8cm}{6cm}
%\epsfxsize=2cm
%\begin{center}
%\begin{figure}
\includegraphics{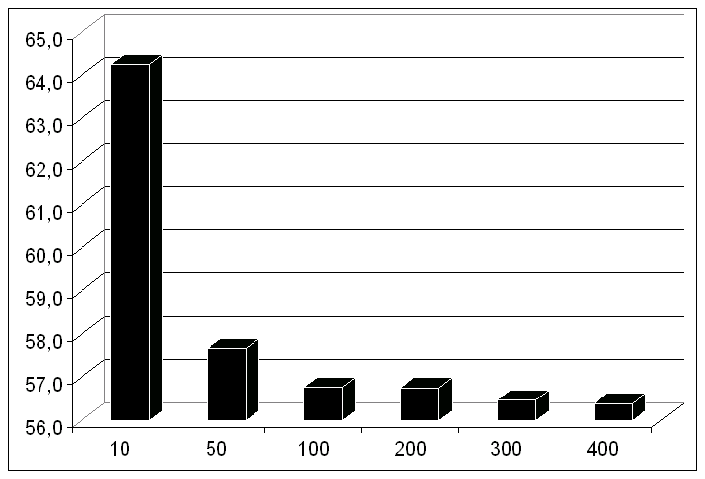}

%\end{center}
\caption{Dependency of $\chi^2$ on number of members in
  each generation}

\end{figure}

\section{Conclusions}
Analysis of QPO's in neutron star atoll sources in the framework of
Hartle--Thorne geometry gives a very detailed data (neutron star
parameters as mass,spin and quadrupole momentum) that could be quite
well used for constraining the wide scale of allowed EOS
constraining the structure of neutron stars by solving the complex
set of structure differential equations. The testing by standard
approaches is a long time consuming procedure. Here, we show in the
case of atoll source 4U 1636--53 that the genetic algorithm method
could make the proper selection in a wide sample of EOS of Skyrmion
type in a very efficient and short way (with same precision and the
time consumed for the GA about 30 minutes, being by orders shorter
that the time consumed by the standard methods). We can conclude
that using the GA with chromosome (EOS,$\rho_c$, $\Omega$) and
putting physically motivated restrictions on the angular velocity
($\Omega$) and central density ($\rho_c$) governing the neutron star
model being an output of the structure equations and the fitness
function $\chi^2$ in the algorithm procedure, we are able to find
the most probably structure of neutron star that fits the observed
QPO data, with respect to observed rotational frequency for the
source 4U 1636--53. It is given by the chromosome ($1,~1.79\cdot
10^{15},~1837.66$).

We expect that the method allows much strongest test
by the genetic algorithm including the quadrupole momentum calculated
directly without using the quasi--Kerr approximation $q\sim j^2$.

\section{Acknowledgement}%ACKNOWLEDGEMENT
This work was supported by Czech grants MSM 4781305903 (Z. S., P. C.
and G. T.) and LC 06014 (M. U. and P. B.).

%\end{minipage}

\end{document}